\documentclass[12pt]{article}
\usepackage[english,german,french,polish]{babel}
\usepackage[T1]{fontenc}
\usepackage{amsfonts}

\begin{document}

\baselineskip 0.75cm
\topmargin -0.6in
\oddsidemargin -0.1in

\let\ni=\noindent

\renewcommand{\thefootnote}{\fnsymbol{footnote}}

\newcommand{\SM}{Standard Model }

\newcommand{\SMo}{Standard-Model }

\pagestyle {plain}

\setcounter{page}{1}

\pagestyle{empty}
~~~

\begin{flushright}
\end{flushright}

\vspace{0.3cm}

{\large\centerline{\bf Overall empirical formula for mass spectra}}
{\large\centerline{\bf of leptons and quarks{\footnote{Work supported in part by Polish MNiSzW research grant N N202 103838 (2010--2012).}} }}

\vspace{0.5cm}

{\centerline {\sc Wojciech Kr\'{o}likowski}}

\vspace{0.3cm}

{\centerline {\it Institute of Theoretical Physics, Faculty of Physics, University of Warsaw }}

{\centerline {\it Ho\.{z}a 69,~~PL--00--681 Warszawa, ~Poland}}

\vspace{1.2cm}

{\centerline{\bf Abstract}}

\vspace{0.3cm}

We present an overall empirical formula that, after specification of its free parameters, describes precisely the mass spectrum of charged leptons and is suggested to reproduce correctly also the mass spectra of neutrinos and up and down quarks (together, twelve masses with eight free parameters are presented). Then, it predicts $m_\tau = 1776.80$ MeV , $m_{\nu_1} \rightarrow 0$ eV and $m_d = 5.0$ MeV , $m_s = 102$ MeV, respectively, when the remaining lepton and quark masses, $m_e$, $m_\mu$, $\Delta m^2_{21} = m^2_{\nu_2}-m^2_{\nu_1}$, $\Delta m^2_{32} = m^2_{\nu_3}-m^2_{\nu_2}$ and $m_u$, $m_c$, $m_t$, $m_b$, are taken as an input.

\vspace{0.6cm}

\ni PACS numbers: 12.15.Ff , 12.90.+b 

\vspace{0.8cm}

\ni January 2012

\vfill\eject

~~~
\pagestyle {plain}

\setcounter{page}{1}

\vspace{0.3cm}

Any triplet of particle masses, as these for leptons and quarks, can be phenomenologically parametrized in very different ways by making use of three free parameters. When some parameters are constrained {\it a priori}, one may get various mass predictions, correct or wrong.

In this note, we will consider the particular parametrization 

\vspace{-0.2cm}

\begin{equation}
m_N  =  \rho_N \,\mu \left[ N^2 + \frac{\varepsilon-1}{N^2}- \eta(N-1)\right] 
\end{equation} 

\ni in terms of three mass-dimensional parameters

\vspace{-0.3cm}

\begin{equation}
\mu\;\;,\;\; \mu\varepsilon \;\;,\;\; \mu \eta \;,
\end{equation}

\ni where

\vspace{-0.3cm}

\begin{equation} 
N = 1\;,\;3\;,\;5
\end{equation}

\ni is a quantum number numerating the masses 

\vspace{0.2cm}

\begin{equation}
m_1\;\;,\;\;m_3\;\;,\;\;m_3\;,
\end{equation}

\ni while

\vspace{-0.2cm}

\begin{equation}
\rho_1 = \frac{1}{29}\;\;,\;\;\rho_3 = \frac{4}{29}\;\;,\;\;\rho_5 = \frac{24}{29}
\end{equation}

\ni stand for generation-weighting factors satisfying the normalization condition $\sum_N \rho_N = 1$. Explicitly, the formula reads

\vspace{-0.2cm}

\begin{eqnarray}
m_1 & = & \frac{\mu}{29} \,\varepsilon  \,, \nonumber \\
m_3 & = & \frac{\mu}{29}\,\frac{4}{9}(80 +\varepsilon - 18\eta)\,, \nonumber \\
m_5 & = & \frac{\mu}{29}\,\frac{24}{25}\,(624 + \varepsilon - 100\eta) \,. 
\end{eqnarray}

\vspace{0.2cm}

\ni It is a transformation of parameters $\mu, \varepsilon, \eta$ into masses $m_1, m_3, m_5$. Its inverse transformation gets the form 

\begin{eqnarray}
\mu & = &  \frac{29}{12928}\left[75m_5-4(225m_3 - 82m_1)\right] \;,\nonumber \\  
\varepsilon & = &  \frac{29}{\mu} m_1 = \frac{12928 m_1}{75m_5-4(225m_3 - 82m_1)}\;, \nonumber \\
\eta & = &  \frac{8}{3}\,  \frac{125 m_5-6(351m_3 - 136m_1)}{75m_5-4(225m_3 - 82m_1)} \;,
\end{eqnarray} 

\ni allowing to fit the free parameters $\mu, \varepsilon, \eta$ to experimental values of mases $m_1, m_3, m_5$. If this can be done with some of the parameters constrained {\it a priori}, we may obtain some predictions for the mass spectrum.

The reason  why we consider here the particular formula (1) is its wonderful propriety of precisely reproducing the triplet of charged lepton masses $m_e, m_\mu, m_\tau$ , when we impose {\it a priori} the constraint

\vspace{-0.2cm}

\begin{equation}
\eta = 0 \,. 
\end{equation}

\ni In fact, for $m_1 = m_e \,,\, m_3 = m_\mu \,,\, m_5 = m_\tau $, the third formula (7) with $\eta = \eta^{(e)} = 0$ gives the prediction [1] 

\begin{equation}
m_\tau = \frac{6}{125} \left(351m_\mu - 136 m_e\right) = 1776.7961\;{\rm MeV} = 1776.80\;{\rm MeV} 
\end{equation}

\ni {\it versus} the experimental value [2] 

\vspace{-0.3cm}

\begin{equation}
m_{\tau}  =  1776.82 \pm 0.16\;{\rm MeV} \,,
\end{equation}

\ni when the experimental values $m_e = 0.5109989$ MeV and $m_\mu = 105.65838 $ MeV [2] are used as the only input. The first two formulae (7) determine then the parameters

\begin{equation}
\mu^{(e)} = \frac{29(9m_\mu-4 m_e)}{12928} = 85.9924 \;{\rm MeV} \;\;,\;\;\varepsilon^{(e)} = \frac{320m_e}{9m_\mu-4 m_e} = 0.17229\,.
\end{equation}

Now, let us try to impose {\it a priori} on the formula (1) or (6) the constraint

\vspace{-0.2cm}

\begin{equation}
\varepsilon \rightarrow 0 
\end{equation}

\ni and conjecture that it is the case for neutrinos: $m_1 = m_{\nu_1}\,,\;m_3 = m_{\nu_2}\,,\;m_5 = m_{\nu_3}$ with $\varepsilon = \varepsilon^{(\nu)} \rightarrow 0 $. In this case, from the second formula (7) we predict

\vspace{-0.2cm}

\begin{equation}
m_{\nu_1} \rightarrow 0 
\end{equation}

\ni and determine then that

\vspace{-0.2cm}

\begin{eqnarray}
m_{\nu_2} & = & \sqrt{\Delta m^2_{21}+ m^2_{\nu_1}}\rightarrow \sqrt{\Delta m^2_{21}}  = 8.8 \times 10^{-3}\; {\rm eV} \;, \nonumber \\
m_{\nu_3} & = & \sqrt{\Delta m^2_{32}+ m^2_{\nu_2}}\rightarrow \sqrt{\Delta m^2_{32}+ \Delta m^2 _{21}}  = 5.0 \times 10^{-2}\; {\rm eV}\;,
\end{eqnarray}

\ni when we use the experimental estimates $\Delta m^2_{21} \equiv m^2_{\nu_2}-m^2_{\nu_1} = 7.7\times 10^{-5}\,{\rm 
eV}^2$ and $\Delta m^2_{32} \equiv m^2_{\nu_3}-m^2_{\nu_2} = 2.4\times 10^{-3}\,{\rm eV}^2$ (assuming that $\Delta m^2_{32}> 0$) [2]. From the formulae (7) we determine the parameters $\mu^{(\nu)} = -0.93 \times 10^{-2}$ eV and $\eta^{(\nu)} = 7.9$ as responsible for $m_{\nu_2} = 8.8\times 10^{-3}\,{\rm eV}$ and $m_{\nu_3} = 5.0\times 10^{-2}\,{\rm eV}$. 

In the case of up and down quarks, the concept of mass loses its observable status getting rather an effective character because of their confinement in hadrons ({\it i.e.}, their nonasymptotic behavior). In this note, we will accept for our discussion the particular quark-mass estimates given in Ref. [2]:

\vspace{-0.2cm}

\begin{eqnarray}
m_{u,d} & = & \left\{\begin{array}{r} 1.7 \;{\rm to}\;3.3\;{\rm MeV} \\ 4.1\; {\rm to}\; 5.8\; {\rm MeV} \end{array} \right.  \rightarrow  \left\{\begin{array}{r} 2.5\;{\rm MeV} \\ 5.0\; {\rm MeV} \end{array} \right. \,, \nonumber \\
m_{c,s} & = & \left\{\begin{array}{r} \!1.27^{+0.7}_{-0.09}\;\; {\rm GeV} \\ \;\; 101^{+29}_{-21}\;\;\;  {\rm MeV}  \end{array} \right. \;,  \nonumber \\
m_{t,b} & = & \left\{\begin{array}{r} \!172.0 \pm 2.2\;{\rm GeV}  \\ 4.19^{+0.18}_{-0.26}\;{\rm GeV} \end{array} \right.  
\end{eqnarray}

\ni (for $m^{(u,d)}$ we take mean values). Then, let us try to fit to these effective masses (to their central values, for simplicity) the free parameters $\mu^{(u,d)}\,,\,\varepsilon^{(u,d)}\,,\,\eta^{(u,d)}$ through the inverse transformations (7), obtaining

\begin{equation}
\mu^{(u,d)} = \left\{\!\!\!\begin{array}{ll} 26.4 \!\!&\!\! {\rm GeV} \\ \,\,\,0.505 \!\!&\!\! {\rm GeV} \end{array}\right.\,,\, 
\varepsilon^{(u,d)} = \left\{\!\!\!\begin{array}{l} \,0.0027  \\ \;0.29 \!\!\end{array} \right.\,,\, 
\eta^{(u,d)} = \left\{\!\!\!\begin{array}{l} \,4.27 = 4 + 0.27  \\ \,3.73 = 4 - 0.27 \!\!\end{array} \right.\,.  
\end{equation}

\ni We can see from Eqs. (16) that

\vspace{-0.2cm}

\begin{equation}
 \frac{2}{29}\mu^{(u)}\varepsilon^{(u)} = \frac{1}{29}\mu^{(d)}\varepsilon^{(d)} = 5.0 \; {\rm MeV} 
\end{equation}

\ni ({\it cf.} $2m_u = m_d = 5.0$ MeV) and 

\vspace{-0.2cm}

\begin{equation}
\eta^{(u)} - 4 = -\eta^{(d)} +4 = 0.27 \,. 
\end{equation}

\ni Let us conjecture that the relations

\vspace{-0.2cm}

\begin{equation}
(\kappa \,\equiv)\, \frac{2}{29} \mu^{(u)}\varepsilon^{(u)} = \frac{1}{29} \mu^{(d)}\varepsilon^{(d)}\;\;,\;\;(\omega \,\equiv)\, \eta^{(u)} - 4 = -\eta^{(d)} +4 
\end{equation}

\ni are valid {\it strictly} (not only approximately as in Eqs. (17) and (18)). We treat them as constraints imposed {\it a priori} on the parameters $\mu^{(u,d)} \varepsilon^{(u,d)}$ and $\eta^{(u,d)}$. Then, we are left with only four free parameters $\mu^{(u,d)} , \kappa$ and $\omega$. Three of them, say $\mu^{(u)} , \kappa$ and $\omega$, can be fitted to the experimental estimates (15) of three masses $m_u , m_c , m_t$, becoming 

\vspace{-0.2cm}

\begin{equation}
\mu^{(u)} = 26.4 \;{\rm GeV} \;,\; \kappa = 5.0\; {\rm MeV}  \;,\; \omega = 0.27 \,.
\end{equation}

\ni Then, the mass $m_d$ is predicted from the first formula (6)

\vspace{-0.2cm}

\begin{equation}
m_d = \frac{\mu^{(d)}}{29}\varepsilon^{(d)} = \kappa = \;5.0 \; {\rm MeV} 
\end{equation}

\ni and subsequently, the mass $m_s$ is also predicted from the third formula (7)  (with $\eta^{(d)} = 4-\omega $) solved with respect to $m_s$: 

\vspace{-0.2cm}

\begin{equation}
 m_s = \frac{25(4+9\omega)m_b + 24(108+41\omega)\kappa}{108(56+25\omega)}= 102 \; {\rm MeV} \,.
\end{equation}

\ni Both masses are in a neat consistency with the experimental estimate (15). The fourth parameter $\mu^{(d)}$ can be now determined from the first formula (7),

\vspace{-0.3cm}

\begin{equation}
\mu^{(d)} = 0.505 \; {\rm GeV}\,, 
\end{equation}

\ni where the experimental estimate (15) of $m_b$ is also used as an input.

Finally, we would like to comment on the possible physical meaning of the quantum number $N = 1,3,5$ appearing in our overall empirical formula (1) for mass spectra of leptons and quarks. It is natural to presume that the quantum field of any fundamental fermion (lepton or quark) should carry an odd number of Dirac bispinor indices ${\alpha_1,\alpha_2,\ldots,\alpha_N}$: $\psi_{\alpha_1 \alpha_2\ldots \alpha_N}(x)$. Among them, one bispinor index, say $\alpha_1$, can be correlated with the \SM $SU(3)\times SU_L(2)\times U(1)$ label (suppressed here) identifying the considered fermion with a lepton or quark. So, this index $\alpha_1$ is distinguished from the remaining bispinor indices, $\alpha_2,\ldots,\alpha_N$ which are, in a natural way, expected to be undistinguishable from each other. Therefore, the bispinor indices ${\alpha_2,\ldots, \alpha_N}$ behave as physical objects obeying Fermi statistics along with Pauli exclusion principle requiring them to be fully antisymmetrized. This implies that $N$ can be equal to 1,3,5 only (since any $\alpha_i$ assumes four values 1,2,3,4), and that the total spin of a fundamental fermion is reduced to spin 1/2 connected with the distinguished bispinor index $\alpha_1$. Hence, we can conclude that in Nature there are exactly three generations of leptons and quarks [3]. 

The fundamental fermions with $N = 1,3,5$ satisfy three Dirac equations that can differ by their mass terms. The gamma matrices in these Dirac equations with $N = 1,3,5$ (fulfilling the Dirac square-root condition $\sqrt{p^2} \rightarrow \Gamma^\mu_N p_\mu$) are [3] 

\vspace{-0.2cm}

\begin{equation}
\Gamma_N^\mu \equiv \frac{1}{\sqrt{N}} \left(\gamma_1^{\mu} + \gamma_2^{\mu} + \ldots + \gamma_N^{\mu} \right) = \gamma^\mu \otimes \underbrace{{\bf 1} \otimes \ldots \otimes {\bf 1}}_{N-1\;{\rm times}} \,,  
\end{equation}

\ni where 

\vspace{-0.3cm}

\begin{equation}
\{\gamma_i^\mu , \gamma_j^\mu\} = 2 g^{\mu\nu} \delta_{ij} \;\;\; (i,j = 1,2,\ldots, N)  
\end{equation}

\ni form a Clifford algebra, while $\{\Gamma_N^\mu , \Gamma_N^\mu\} = 2 g^{\mu\nu} \underbrace{{\bf 1}\otimes {\bf 1}\otimes \ldots \otimes{\bf 1}}_{N\,{\rm times}}$ and $\{\gamma^\mu , \gamma^\nu\} = 2 g^{\mu\nu}$ define Dirac algebras, the second of them being the familiar one ($\gamma^\mu $ and {\bf 1} are conventional $4\times 4$ Dirac matrices).

Note in addition that, in this model, the form (5) of fermion generation-weighting factors turns out to be justified [3]. This is due to the fact that only one component $\psi_{\alpha_1}$ of the field $\psi_{\alpha_1 \alpha_2\ldots \alpha_N}(x)$ for $N =1$, its four components $\,\psi_{\alpha_1 12}=-\psi_{\alpha_1 21} = \psi_{\alpha_1 34}=-\psi_{\alpha_1 43}$ for $N=3$ and twenty four components being permutations of $\psi_{\alpha_1 1234}$ (equal to each other up to sign) for $N=5$ can be different from zero, when our "intrinsic Pauli principle"~as well as special relativity and probabilistic interpretation of quantum theory [3] are invoked (then, for $N=3$ twelve components $\psi_{\alpha_1 14}=-\psi_{\alpha_1 41} ,\, \psi_{\alpha_1 32}=-\psi_{\alpha_1 23} ,\,\psi_{\alpha_1 13}=-\psi_{\alpha_1 31} ,\, \psi_{\alpha_1 24}=-\psi_{\alpha_1 42}$ and $\psi_{\alpha_1 11} ,\, \psi_{\alpha_1 33} ,\, \psi_{\alpha_1 22},\,\psi_{\alpha_1 44}$ are zero). Here, the chiral representation is used, where $1 = (\uparrow,+1) ,\, 2= (\downarrow,+1) ,\,3= (\uparrow,-1) ,\, 4= (\downarrow,-1)$ with $\uparrow \downarrow$ being spin-1/2 projections and $\pm 1$ eigenvalues of chirality $\gamma^5$.

In this way, we can construct an "intrinsically composite model"~ of leptons and quarks of three generations [3], where  Dirac bispinor indices  play the role of "intrinsic partons"~ of which all but one are undistinguishable and obey the "intrinsic Pauli principle", in contrast to one of them which is distinguished by "$\!\,$carrying"~the \SMo label of a lepton or quark. In the present note, we propose in this context a possible overall empirical formula for the mass spectra of leptons and quarks, where $N= 1,3,5$ is a quantum number.

Our intrinsically composite formalism outlined above might either have a fundamental character or be a dynamical 
$S$-wave approximation to a more conventional composite model for an odd number of spin-1/2 orbital partons bound mainly in $S$-states, where one of these partons is distinguished from the rest. In the first option, the conventional orbital partons appearing in the second option are replaced through an act of algebraic abstraction by new algebraic partons described by Dirac bispinor indices. Somewhat similarly, the fundamental notion of Dirac spin 1/2 arises from the notion of orbital angular momentum through an act of algebraic (group-theoretical) abstraction.

~~~~
\vspace{0.5cm}

{\centerline{\bf References}}

\vspace{0.5cm}

{\everypar={\hangindent=0.6truecm}
\parindent=0pt\frenchspacing

{\everypar={\hangindent=0.6truecm}
\parindent=0pt\frenchspacing

~[1]~W.~Kr\'{o}likowski, {\it Acta Phys. Polon.} {\bf B 33}, 2559 (2002); {\it Acta Phys. Polon.}{\bf B 41}, 649 (2010); arXiv: 1009.2388 [{\tt hep--ph}]}; arXiv: 1011.1120 [{\tt physics, gen--ph}]; and references therein. 

\vspace{0.2cm}

~[2]~N.~Nakamura {\it et al.} (Particle Data Group), {\it J. Phys}, {\bf G 37}, 075021 (2010).

\vspace{0.2cm}

~[3]~W.~Kr\'{o}likowski, {\it Acta Phys. Polon.} {\bf B 23}, 933 (1992); {\it Phys. Rev.}, {\bf D 45}, 3222 (1992); {\it Acta Phys. Polon.}, {\bf B 33}, 2559 (2002).

\vspace{0.2cm}

\vfill\eject

\end{document}